\documentclass[runningheads]{svmult}

\usepackage{makeidx}   % allows index generation
\usepackage{graphicx}  % standard LaTeX graphics tool
\usepackage{subeqnar}  % subnumbers individual equations
\usepackage{multicol}  % used for the two-column index
\usepackage{physprbb}  % modified textarea for proceedings,
\makeindex             % used for the subject index
\newcommand{\mathM}{\hbox{$\mathcal{M}$}}

\def\lesssim{\mathrel{\hbox{\rlap{\hbox{\lower4pt\hbox{$\sim$}}}\hbox{$<$}}}}
\def\gtrsim{\mathrel{\hbox{\rlap{\hbox{\lower4pt\hbox{$\sim$}}}\hbox{$>$}}}}

\begin{document}
\title*{The Great Observatories Origins Deep Survey}
\toctitle{The Great Observatories Origins Deep Survey}

\titlerunning{The Great Observatories Origins Deep Survey}

\author{Mark Dickinson\inst{1}
\and Mauro Giavalisco\inst{1}
\and the GOODS team\inst{2}}
\authorrunning{Mark Dickinson}

\institute{Space Telescope Science Institute, Baltimore MD 21218, USA
\and STScI/ESO/ST-ECF/JPL/SSC/Gemini/U.~Fla./Yale/Boston~U./U.~Ariz./IAP/\\
Saclay/UCLA/UC~Berkeley/UCSC/U.~Hawaii/U.~Wisc./PSU/Caltech/GSFC/AUI}

\maketitle              % typesets the title of the contribution

\begin{abstract}

The Great Observatories Origins Deep Survey (GOODS) is designed to
gather the best and deepest multiwavelength data for studying the 
formation and evolution of galaxies and active galactic nuclei, 
the distribution of dark and luminous matter at high redshift, the 
cosmological parameters from distant supernovae, and the 
extragalactic background light.   The program uses the most powerful 
space-- and ground--based telescopes to cover two fields, each 
$10^\prime \times 16^\prime$, centered on the Hubble Deep Field North
and the Chandra Deep Field South, already the sites of extensive 
observations from X--ray through radio wavelengths.  GOODS incorporates 
3.6--24$\mu$m observations from a {\it SIRTF} Legacy Program, 
four--band ACS imaging from an {\it HST} Treasury Program, 
and extensive new ground--based imaging and spectroscopy.
GOODS data products will be made available on a rapid time--scale, 
enabling community research on a wide variety of topics.  
Here we describe the project, emphasizing its application for 
studying the mass assembly history of galaxies.

\end{abstract}

\section{Introduction}

This conference, and this volume of contributions, demonstrate the
vital interest in understanding the mass assembly history of galaxies.
Theory provides guidance about how dark matter halos are
built up in a hierarchical process largely controlled by the 
power spectrum of density fluctuations and the parameters of 
the cosmological world model.  The assembly of the stellar content 
of galaxies is governed by more complex physics, including gaseous 
dissipation, the mechanics of star formation itself, and feedback due
to the energetic output from stars and AGN on the baryonic material 
within galaxies.

Deep imaging and spectroscopic surveys now routinely find and study 
galaxies throughout most of cosmic history, back to redshifts 
$z = 6$ and earlier.   However, observations are only now beginning 
to provide constraints on galaxy mass assembly, particularly at $z > 1$.
For example, the stellar mass assembly history is characterized by the evolving 
distribution of masses ($\mathM$) and star formation rates (SFR, or 
$\dot{\mathM}$) with time or redshift, $f(\mathM, \dot{\mathM}, t$). 
Most investigations to date have considered only moments over this distribution, 
such as luminosity functions (an imperfect surrogate for the mass distribution)
or the global star formation rate SFR($z$).  Moreover, at high redshift, 
$\mathM$ and $\dot{\mathM}$ are at best only imperfectly measured 
using currently available observables.   Starlight traces stellar mass only 
in an indirect manner:  the mass--to--light ratio ($\mathM/L$) of a mixed 
stellar population depends on many parameters, including its age, past star 
formation history, initial mass function (IMF), dust extinction, and metallicity.
Locally, the best constraints come from measurements at near--infrared wavelengths 
\cite{gav96,col01}, where the longer--lived stars which dominate the mass 
contribute most to the galaxy luminosity.  Moreover, the effect of dust 
extinction is smaller at redder wavelengths.  For $\dot{\mathM}$, no one
observable provides a direct and ``universal'' tracer of star formation 
in all circumstances.  Ultraviolet, mid-- and far--infrared, radio, 
and nebular line emission are all valuable tools for measuring
star formation, with different dependences on extinction, IMF, etc., 
and thorough surveys of high redshift star formation require the use 
and cross--calibration of multiple indicators.

\section{The Great Observatories Origins Deep Survey}

The Hubble Deep Fields (HDF--N and HDF--S \cite{wil96,wil00,fer00}) provided 
an invaluable resource of public data for studying faint, distant galaxies.  
Moreover, they served as a catalyst for follow--up observations 
at many wavelengths using the most powerful telescope facilities
in space and on the ground.  However, the HDFs have their limitations.
First, they are very small fields, 5~arcmin$^2$ each, probing very small
co--moving volumes.  Second, the HDF--S followed the HDF--N by several 
years.  This diluted its impact somewhat, and reduced motivation for 
the vital follow--up studies needed to verify HDF--N results and to test
their robustness against line--of--sight variations due to galaxy 
clustering.  Third, the wavelength range $\lambda\lambda 3$--1000$\mu$m
is the ``weak link'' in HDF coverage.  ISO data at 7 and 15$\mu$m 
\cite{ser97,oli02} and SCUBA measurements at 850$\mu$m \cite{hug98} probe 
mid-- and far--IR emission, but can detect only the most luminous 
dust--obscured objects at high redshift.  HDF studies of galaxies at 
$z > 1$ are therefore missing important information at mid-- and 
far--infrared wavelengths where most of the bolometric luminosity from 
star formation is believed to emerge, as well as the redshifted 
near--infrared rest--frame light ($\lambda\lambda 3$--$10\mu$m) 
which most nearly traces total stellar mass.

The Great Observatories Origins Deep Survey (GOODS) follows in the footsteps 
of the HDF projects, and is a campaign to unite the best, deepest 
data across the electromagnetic spectrum to create a community resource for 
exploring the distant universe.   GOODS data will be used to study the formation 
and evolution of galaxies, the radiative output from active galactic nuclei 
and star formation at high redshift, the characteristics of the extragalactic 
background light, large scale structure and the distribution of 
dark matter, the values of the cosmological parameters, and many other 
projects outside the scope of its core design.

GOODS builds upon existing or ongoing surveys from space-- 
and ground--based facilities, including NASA's Great Observatories, 
{\it HST}, {\it Chandra} and {\it SIRTF}.  The program targets two
fields, each $10^\prime \times 16^\prime$, around the Hubble Deep Field North 
(HDF--N) and the Chandra Deep Field South (CDF--S).  These are the most 
data--rich and well--studied deep survey areas on the sky, with extensive 
near--infrared and optical imaging and spectroscopy, highly sensitive radio 
and sub--mm measurements, and the deepest X--ray observations from 
{\it Chandra} \cite{gia02,bra01} and {\it XMM--Newton} (in progress; 
PIs: Bergeron (CDF--S); Jansen and Griffiths (HDF--N)).  Two fields, one 
in each celestial hemisphere, provide insurance against variance due to 
line--of--sight clustering effects, and enable follow--up programs by 
astronomers and observatories worldwide.

\subsection{The {\it SIRTF}\, Legacy Program}

\begin{figure}[t]
\begin{center}
\includegraphics[width=\textwidth]{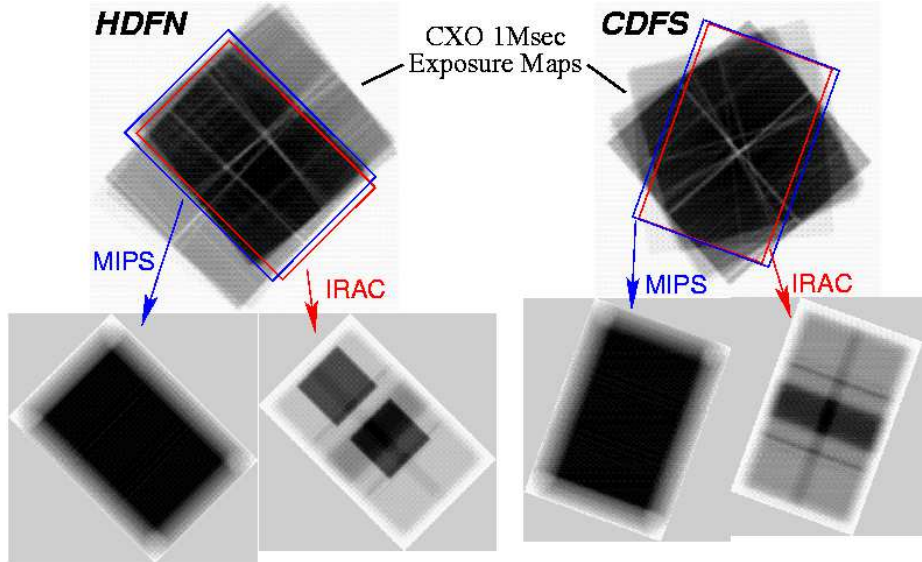}
\end{center}
\caption[]{Layout of the GOODS/{\it SIRTF} fields.  The upper 
panels show the exposure maps for the 1~Msec {\it Chandra} X--ray 
observations of the HDF--N and CDF--S.   Boxes show the approximate 
boundaries of the {\it SIRTF} IRAC and MIPS survey areas 
(roughly $10^\prime \times 16^\prime$).  Insets below
show the nominal {\it SIRTF} exposure time maps.  The HDF--N
includes ``ultradeep'' IRAC observations, which are contingent upon
on--orbit tests to establish the practical sensitivity limit of
the instrument.}
\label{fig1}
\end{figure}

The GOODS {\it SIRTF} Legacy Program (PI: Dickinson) will make the 
deepest observations with that facility at 3.6 to 24$\mu$m.  Observations
will be carried out in the first year of {\it SIRTF} operations, in 
2003--2004.  The bulk of the GOODS {\it SIRTF} program will use the 
Infrared Array Camera (IRAC), observing at 3.6, 4.5, 5.8 and 
8.0$\mu$m with exposure times of 23.6 hours per band.   A small overlap 
strip in each field will receive twice this integration time, and in 
the HDF--N only, a pair of $5^\prime \times 5^\prime$ ``ultradeep'' IRAC 
fields are planned with exposure times of 70 hours (reaching
94 hours in the maximum overlap region atop the WFPC2 HDF--N).

These long exposure times are essential in order to reach sensitivities
$\lesssim 1\mu$Jy with reasonably high S/N ratios.  For example, the IRAC 
8$\mu$m observations will sample the rest--frame $K$--band light 
from Lyman break galaxies (LBGs) at $z \approx 3$, and will thus 
provide an important handle on their total stellar content.
The expected 8.0$\mu$m flux for an ``$L^\ast$'' LBG 
(with $\mathM_\ast \approx 10^{10} \mathM_\odot$ \cite{pap01,sha01}) 
is $1.5 \mu$Jy.  At 3.6 and 4.5$\mu$m, the flux sensitivity achieved 
by the GOODS/IRAC observations will depend strongly on the achieved 
image quality (expected to be in the range 1.5--2.3 arcsec FWHM),
since source confusion will be important -- less so at 5.8 and 
8.0$\mu$m, where the zodiacal background should set the flux limits.  
The ultradeep IRAC fields will probe farther down
the luminosity and mass function at $z \sim 3$, and should detect
typical objects at $z \approx 5$.

The GOODS fields will also be observed at longer wavelengths with
the {\it SIRTF}/MIPS instrument.  Deep ISOCAM 15$\mu$m imaging surveys 
were sensitive to redshifted $7.7\mu$m PAH emission from star--forming 
galaxies at $z \approx 1$, and the GOODS 24$\mu$m observations are 
designed to detect objects with similar rest--frame luminosities at 
$z = 2$ to 2.5.  In principle, they should be able to detect the 
mid--infrared emission from obscured star formation in typical Lyman 
break galaxies at these redshifts.  The actual sensitivity achieved 
at 24$\mu$m will depend on the (presently uncertain) level of source 
confusion and on instrument performance.  The GOODS program
plans 10.4 hour exposures at 24$\mu$m, contingent upon on--orbit 
demonstration that they will reach substantially fainter flux limits 
than planned 20~min exposures from the MIPS GTO wide--field survey 
(PI: Rieke) which covers the GOODS fields.  Test observations made early 
in the mission will be used to determine the longest exposure times 
practical for making confusion--limited observations of the GOODS fields.
The {\it SIRTF} GTO program will also cover these fields at 70$\mu$m 
and 160$\mu$m, sensitive to the far--infrared thermal emission from 
high redshift, dust--obscured star formation.

\subsection{The {\it HST}/ACS Treasury Program}

The GOODS {\it HST} Treasury Program (PI: Giavalisco) will use the 
Advanced Camera for Surveys (ACS) to image the fields with four broad,
non--overlapping filters, F435W ($B$), F606W ($V$), F775W ($i$), and 
F850LP ($z$), with exposure times of 3, 2.5, 2.5 and 5 orbits, respectively,
reaching extended--source sensitivities within 0.5-0.8 mags of the 
WFPC2 HDF observations.  The observations will be carried out during 
{\it HST} Cycle 11, in 2002--2003.  GOODS is a deep survey, not a wide one, 
but is nevertheless much larger than most previous {\it HST}/WFPC2 programs.
The GOODS fields cover $32\times$ the solid angle of the combined HDF-N and S, 
and are $4\times$ larger than the combined HDF Flanking Fields, and 
$2.5\times$ larger than the WFPC2 Groth Strip Survey.  The $Viz$ observations 
will be taken in five repeat visits separated by approximately 45 days, 
enabling a search for SNe~Ia at $1.2 < z < 1.8$ to test the apparent 
transition from cosmic deceleration to acceleration that is
predicted in world models dominated by a cosmological constant 
(and suggested by current data \cite{rie01}).  

The $z$--band observations will image the optical rest--frame light from 
galaxies out to $z = 1.2$, with angular resolution superior to that from 
WFPC2.   The $BViz$ imaging will also enable a systematic survey of 
Lyman break galaxies at $4 < z < 6.5$, reaching back to the suggested epoch 
of reionization \cite{bec01,djo01}.   The photometric depth and co--moving 
volume coverage will make it possible to quantify the LBG population in this 
redshift range with statistical accuracy comparable to that now available from 
large, ground--based LBG surveys at $z \approx 3$~\cite{ste99}.  The ACS data 
will also provide a powerful tool for studies of gravitational lensing, 
low--mass stars in our galaxy, and perhaps objects in the outer solar system.

\begin{figure}[t]
\begin{center}
\includegraphics[width=\textwidth]{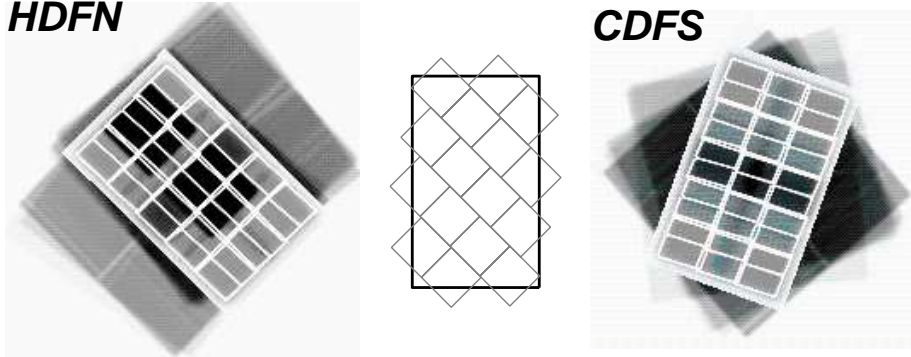}
\end{center}
\caption[]{Layout of the GOODS/{\it HST} observations.
The grid of white boxes shows the tiling of {\it HST}/ACS fields
at one telescope orientation, superimposed on the {\it Chandra} 
(outer greyscale) and {\it SIRTF} IRAC (inner greyscale) exposure 
maps.  The fields will be revisited approximately every 45 days
to enable a search for high redshift supernovae.  The center inset
schematically shows how the rotated ACS pointings from alternate
visits will be tiled over the GOODS area.
}
\label{fig2}
\end{figure}

\subsection{Ground--based observations}

As noted above, the HDF--N and CDF--S are already among the most 
data--rich deep survey regions on the sky, and the GOODS program includes 
a large component of ground--based supporting observations to enable 
research on distant objects.  In large, coordinated NOAO and ESO programs
(PIs: Cesarsky, Dickinson), we are obtaining new optical and near--infrared 
imaging, including $U$--band imaging from the KPNO and CTIO 4m MOSAIC 
cameras, and $JHK_s$ imaging with the KPNO 4m/FLAMINGOS and VLT/ISAAC 
instruments.  We are also planning a spectroscopic campaign in the 
CDF--S using the new VIMOS and red--upgraded FORS--2 spectrographs on 
the VLT (see the contribution by Renzini et al.\ to these proceedings).  
This program will provide a public data resource of several thousand
spectra and redshifts for galaxies in the southern GOODS field.
We will also supplement the already--extensive HDF--N redshift data \cite{coh00} 
and extend it over the whole GOODS/HDF--N using spectroscopy with Gemini--N/GMOS 
and Keck/DEIMOS+LRIS.  We are collaborating on new ATCA radio observations 
of the GOODS/CDF--S (PI: Koekemoer), and JCMT/SCUBA observations of the 
GOODS/HDF--N (PIs: Barger, Scott).  Table~1 provides a summary of observations 
being taken as part of the GOODS project, along with some other key data 
sets for the GOODS fields at other wavelengths.

\begin{table}
\caption{GOODS observations and complementary data sets.  The top portion 
of the table lists GOODS space-- and ground--based imaging observations 
at 0.36--24$\mu$m and their nominal sensitivities.  The bottom portion
is an incomplete list of additional observations available, 
in progress, or in preparation, which cover the GOODS areas.}
\begin{center}
\renewcommand{\arraystretch}{1.4}
\setlength\tabcolsep{5pt}
\begin{tabular}{lll}
\hline\noalign{\smallskip}
Wavelength & Facility & Sensitivity (S/N=5) \\
\noalign{\smallskip}
\hline
\noalign{\smallskip}

0.36$\mu$m	& KPNO+CTIO 4m		& AB = 27.3  ($U$) \\
0.4-0.9$\mu$m	& {\it HST}/ACS		& AB = 27.9, 28.2, 27.5, 27.4$^{\mathrm a}$ 
					    ($BViz$)\\
1.2-2.2$\mu$m	& VLT, KPNO 4m		& AB = 25.2, 24.7, 24.4 ($JHK_s$) \\
3.6-8.0$\mu$m	& {\it SIRTF}/IRAC	& AB = 24.5, 24.5, 23.8, 23.7 
					  (0.6--1.2 $\mu$Jy)$^{\mathrm b}$ \\
24$\mu$m	& {\it SIRTF}/MIPS	& 20-80 $\mu$Jy$^{\mathrm c}$ \\

\noalign{\smallskip}
\hline\noalign{\smallskip}
Type & Facility & Notes \\
\noalign{\smallskip}
\hline
\noalign{\smallskip}

Spectroscopy	& VLT, Gemini, Keck  	&  Various PIs; GOODS programs \& collabs. \\
X-ray		& {\it Chandra}, {\it XMM}  &  Public {\it Chandra} data and {\it
XMM} GTO progs.\\
70, 160$\mu$m	& {\it SIRTF}/MIPS  	&  {\it SIRTF} GTO program \\
Sub-mm		& SCUBA, SEST		&  Various PI programs \\
Radio		& VLA, ATCA		&  Various PIs; CDF--S observs.\ in progress \\

\noalign{\smallskip}
\hline
\end{tabular}
\end{center}
$^{\mathrm a}$ For 0.5 arcsec diameter aperture \\
$^{\mathrm b}$ IRAC deep survey, for ``handbook'' PSF;  
3.6 and 4.5$\mu$m performance may be better \\
$^{\mathrm c}$ Uncertain sensitivity; depends on instrument performance and source 
confusion \\
\label{Table1}
\end{table}

\subsection{Data products}

In the spirit of the HDF projects, the GOODS team will make data products 
available to the community on a rapid time--scale.  The raw {\it SIRTF} and 
{\it HST} data will be available upon ingestion into the SSC and STScI
archives.  The reduced data products from both facilities will be provided 
in a series of incremental releases:  ``best effort'' (version 0.5) reduced 
images two to three months after each observing epoch;  improved (version 1) 
image mosaics three months after the final observations; and reprocessed (version 2) 
data products and multiwavelength catalogs six to twelve months after the final 
observations.  Similar release schedules apply to the ancillary 
data from ESO and NOAO, and we will generally follow similar procedures 
for GOODS--related data from other facilities.

\section{Science enabled by GOODS}

A primary goal of the GOODS program is to provide observational 
data for tracing the mass assembly history of galaxies throughout most 
of cosmic history.  First, this requires redshift information to sort
galaxies by distance and cosmic time.  Much of this will come from
the existing and planned spectroscopic surveys of these fields.  
At fainter fluxes, the 13--band GOODS imaging data, covering 4.5 wavelength 
octaves from 0.36--8$\mu$m, will provide an exceptional resource for estimating 
photometric redshifts for galaxies of all types, calibrated by the extensive 
spectroscopy.

The {\it SIRTF}\, IRAC data is designed to measure the rest--frame $K$--band 
starlight from ``ordinary'' galaxies (e.g., the progenitor fragments
of the Milky Way) at $z \approx 3$, and can detect rest--frame near--infrared 
light ($\lambda > 1\mu$m) from objects out to $z = 7$.  Photometry 
will trace the spectral energy distributions of galaxies from UV 
through IR rest--frame wavelengths, and thus constrain their stellar 
populations and $\mathM/L$, providing the best estimates (modulo assumptions 
about the IMF) of their total stellar masses.  GOODS data, as well as other
observing programs covering these fields, will offer a wide array of star 
formation indicators, including rest--frame UV and mid--infrared 
photometry, far--infrared measurements from the {\it SIRTF} 
GTO MIPS program, very deep radio and sub--mm surveys, and nebular line 
spectroscopy from the redshift surveys and targeted follow--up programs.  
These different indicators can be applied and cross--calibrated for the 
{\it same} high redshift galaxies, guided by detailed knowledge from galaxy 
surveys in the local universe, such as the SINGS {\it SIRTF} Legacy Program 
(PI: Kennicutt).

The {\it HST}/ACS program will provide high resolution imaging needed 
to relate the morphological properties of galaxies (size, surface brightness, 
Hubble type, etc.) to their stellar populations, masses, star formation 
rates, and AGN activity, tracing the emergence of the Hubble sequence 
and its relation to the physical characteristics of galaxy assembly.
Future programs of high--dispersion spectroscopy can be used to measure 
galaxy kinematics which trace gravitational mass, connecting the stellar 
population properties traced by {\it SIRTF} and the dark matter potential 
wells.  Weak and strong lensing measurements, as well as galaxy clustering, 
will also provide statistical constraints on dark halo mass on larger 
physical scales.

GOODS data will also provide an important resource for studying 
the evolution of active galactic nuclei, in particular to identify
both obscured and unobscured AGN with ``typical'' luminosities
(i.e., not just the most powerful QSOs and radio galaxies) out to 
high redshifts, back to the ``QSO era'' at $z > 2$.  Deep X--ray 
data will sort AGN from starbursts as the engines powering mid-- 
and far--IR emission in distant objects, enabling a census of energetic 
output from these mechanisms over a broad range of redshift.
The {\it SIRTF} data will also fill an important gap in our measurements 
of the discrete source component of the extragalactic background light 
(EBL), the integral record of emission and absorption of radiation 
throughout cosmic history.  IRAC data at 3.6--8$\mu$m will trace the 
``downside'' of the peak of the direct stellar contribution to the EBL, 
while 24$\mu$m measurements (and GTO {\it SIRTF} data at 70 and 160$\mu$m) 
will follow the ``upside'' of the far--infrared peak due to dust--absorbed 
starlight and AGN emission.

\section{Conclusion}

The installation of ACS on board {\it HST}, the launch of {\it SIRTF},
and the implementation of a new generation of massively multiplexed 
spectrographs on large ground--based telescopes will open rich new 
opportunities for observing galaxy formation and evolution, out to
very high redshift and early cosmic epochs.
The GOODS project is designed to bring all of these tools to bear on 
common deep survey fields, uniting the best observations at all accessible 
wavelengths. These data will be gathered into a coherent archive for 
public release, enabling community research on a wide variety of topics.
We hope that at future meetings such as this one, GOODS data
will have helped to advance our understanding of galaxy masses
at high redshift, and of the history of galaxy assembly.

More information about the project and observing program
can be found on the GOODS web sites 
at STScI ({\tt http://www.stsci.edu/science/goods}) and 
ESO ({\tt http://www.eso.org/science/goods}).

\vspace{0.2cm}

The authors wish to thank the organizers for hosting 
a wonderful meeting in a magical location (a former insane asylum,
perhaps not coincidentally), and for generous travel support.  
Additional support for the GOODS {\it SIRTF} Legacy Program is 
provided by JPL contract 1224666.

%INDEX%%%%%%%%%%%%%%%%%%%%%%%%%%%%%%%%%%%%%%%%%%%%%%%%%%%%%%%%%%%%%%%
% Please check with the editor of your book whether he plans to
% include a "mutual" subject index - if so, please code your entries
% in the standard syntax. For your own purposes you may print your
% "personal" index by using the following commands:
%
%\clearpage
%\addcontentsline{toc}{section}{Index}
%\flushbottom
%\printindex
%%%%%%%%%%%%%%%%%%%%%%%%%%%%%%%%%%%%%%%%%%%%%%%%%%%%%%%%%%%%%%%%%%%%%

\end{document}